\documentclass[epsf,twocolumn,prb,preprintnumbers,amsmath,amssymb,floatfix,aps,graphicx, showpacs]{revtex4} 
\usepackage{graphicx}

\usepackage{subfigure,bm,dcolumn}

\newcommand{\be}{\begin{equation}}
\newcommand{\ee}{\end{equation}}


\begin{document}

\title{Linear response separation of a solid into atomic constituents:\\
         Li, Al, and their evolution under pressure}
\author{J. Kunstmann,$^{\dag}$ L. Boeri,$^{\dag}$ and W. E. Pickett$^{\dag}$$^{\ddag}$}
\affiliation{$^{\dag}$Max-Planck-Institut f\"ur Festk\"orperforschung, Heisenbergstra{\ss}e 1, 70569 Stuttgart, Germany}
\affiliation{$^{\ddag}$Department of Physics, University of California Davis, Davis, California, 95616, USA}
\date{\today}
\begin{abstract}
We present the first realization of the generalized
 pseudoatom concept introduced by Ball, and adopt the name {\em enatom} to 
minimize confusion.
This enatom, which consists of a unique 
decomposition of the total charge density
(or potential) of any solid into a sum of overlapping atomiclike contributions that
move rigidly with the nuclei to first order, is calculated using (numerical) linear
response methods, and is analyzed for both fcc Li and Al at pressures of 0, 35, and 50 GPa.
These two simple fcc metals (Li is fcc and a good superconductor in the 20-40 GPa range) show
different physical behaviors under pressure, which
reflects the increasing covalency in Li and the lack of it in Al.  The nonrigid 
(deformation) parts of the enatom charge and potential have opposite signs in Li 
and Al; they become larger under pressure only in Li.
These results establish a method of construction of the enatom, whose potential can be used to obtain a real-space understanding of the vibrational properties and electron-phonon interaction in solids.

\end{abstract}
\maketitle

\section{Introduction}
The desire to describe condensed matter as a collection of (generalized)
atoms is older than the discipline of solid state physics itself, extending
back to the ancient Greek atomists.\cite{demo,book}  This passion remains 
as strong as ever, embodied in site-based models, atomic orbital calculational
approaches, and in a
resurgence of interest in Wannier functions.  At the other extreme lies
the homogeneous electron gas starting point for condensed systems,
where the atomic nature is disregarded at first and brought in only
as necessary to address the realities of real solids.  These two 
viewpoints comprise the traditional viewpoints of solid state chemistry
and condensed matter physics, respectively.
The simple model of overlapping free atoms is useful pedagogically but neglects
the real physics of the formation of solids.

Anti-intuitively, the itinerant weak-pseudopotential
viewpoint provided the first example of how
the density $n(\bm r)$ and the electronic potential $v(\bm r)$ can be 
separated uniquely into atomic contributions. 
In the limit of weak pseudopotentials, the {\it neutral pseudoatom} approach of
Ziman\cite{ziman} gives a description of a nearly-free-electron solid
as overlapping atomic contributions, a description that extends to
lattice dynamics and even to the melt.  It is however severely restricted
by the limitation to weak pseudopotentials, which applies
only to the alkali metals and may not give satisfactory accuracy even
there.  

Dagens defined an {\it auxiliary neutral atom} arising from a change in
density (with zero net charge) induced by a screened potential in jellium.\cite{dagens}
This density functional theory inspired prescription was solved numerically
for Li and Na, but does not address the overlapping of such entities.
In the work of Streitenberger \cite{streitenberger} a \textit{generalized pseudoatom model} is introduced that extends the pseudoatom concept of Ziman to inhomogeneous  electron-ion systems for simple metals.
The model is based on linear-response theory in the density functional framework
and is applied to a metal surface.


\subsection{\label{sec:background}Formal background}

Three decades ago Ball introduced a \textit{generalized pseudoatom}  
density decomposition concept, one that applies to any solid. \cite{Ball1,Ball2}
It is this specification that we follow in this paper.
The development of the broad pseudoatom concept and
the terminology (pseudoatom; auxiliary neutral atom; generalized pseudoatom; 
quasi-atom) has a long history, and the term pseudoatom also means `atom described by a pseudopotential' and `phantom atom to tie off dangling bonds', as well as many other applications, as a literature search will readily reveal.
It will therefore be useful to introduce unambiguous language in the following:
instead of using the \textit{generalized pseudoatom} terminology of Ball we 
introduce the term {\it enatom}. 
\footnote{Greek: "en" denotes within or inside; "atom" denotes indivisible
part. Therefore "enatom" connotes the indivisible part inside
a system.}
It should be clarified that Ball's pseudoatom has nothing to do with any
pseudopotential (a common use of the term).

For any reference position of atoms, the (vector) first-order change in charge density  upon displacing one atom at $\bm R_j$ from its equilibrium position $\bm R^0_j$, i.e., 
the linear response to displacement, can be separated into its irrotational and divergenceless
components
\begin{eqnarray}
\label{eqn:foc}
\frac{\partial n(\bm r)}{\partial \bm R_j}
    &\equiv&\nabla_j n(\bm r) \\ \nonumber
  &=& -\nabla \rho_j(\bm r - \bm R_j^o)
              + \nabla \times \bm B_j(\bm r - \bm R_j^o);
\end{eqnarray}
here $n(\bm r)$ is the charge density of the system.
An immediate result is a pair of remarkable sum rules. \cite{Ball1} 
(i) The lattice sum of the {\it rigid density}\footnote{Our definition of the 
rigid part $\rho$ is predominantly positive [like $n(\bm r)$] and differs in
sign from the convention of Ball.\cite{Ball1}}
$\rho_j(\bm r-\bm R_j^o)$ gives an exact decomposition of the crystal
charge density into atomic contributions:
\be
\label{eqn:sumrule1}
 \sum_j \rho_j(\bm r - \bm R_j^o) = n(\bm r).
\ee
(ii) The lattice sum of the \textit{deformation density}
 (or ``backflow'') $\nabla \times \bm B_j(\bm r - \bm R_j^o)$ 
vanishes identically:
\be
\label{eqn:sumrule2}
   \sum_j \nabla \times \bm B_j (\bm r - \bm R_j^o) = \bm{0}.
\ee
This strong constraint reflects that this nonrigid density is a
cooperative effect of neighboring atoms, which nevertheless can be broken
down uniquely into individual atomic contributions. 
Clearly atoms that are equivalent by symmetry have identical 
$\rho_j$ and $\bm B_j$; an elemental solid with one atom per primitive 
cell only has one of each.

This cooperative origin (a solid state effect)
of the deformation density can be understood 
by considering that an atom, embedded in a jellium
background, would have no deformation density. In fact,
if the density is $n(\bm r)$ when the atom is at the origin, then by 
translational symmetry the density is $n(\bm r - \bm R_j)$ when the atom 
is located at $\bm R_j$.  Therefore $\nabla_j n(\bm r) = -\nabla \rho_j(\bm r)$
and there is no nonrigid contribution. Furthermore, within standard treatments
(such as the local density approximation) $\rho_j(\bm r)$
is spherical, because the atom is embedded in an isotropic environment.  
Thus the deformation density and the anisotropy of the rigid density arise 
solely from the inhomogeneity of the system, i.e., by neighboring atoms.
The other extreme is represented by a covalent solid, which is held together
by strong directional bonds. 
The rigid density will be highly non-spherical, and the
deformation density will be comparatively large, reflecting the fact that when an atom
moves its bonding is disrupted.

To first order in displacements from the reference point $\delta \bm R_j = \bm R_j - \bm R_j^o $
(typically
the equilibrium lattice), the density is given by \cite{Ball1}
\begin{eqnarray}
\label{eqn:dens}
n(\bm r;\{\bm R_j\})&=&\sum_j[\rho_j(\bm r- \bm R_j^o -\delta \bm R_j) \\ \nonumber
            & &+\delta \bm R_j \cdot 
  \nabla \times \bm B_j(\bm r- \bm R_j^o)].
\end{eqnarray}
The quantity inside the sum is the {\it enatom} of atom $j$ and moves rigidly with the nucleus
to first order.  (The $\vec R_j^o$ in the argument of $\vec B_j$ can be replaced
by $\vec R_j$ without changing the expression to first order.)
Analogous decomposition and sum rules apply to the potential $v(\bm r)$:\cite{wep}
\begin{eqnarray}
\label{eqn:pot}
v(\bm r;\{\bm R_j\})&=&\sum_j[V_j(\bm r-\bm R_j^o -\delta \bm R_j) \\ \nonumber
            & &+\delta \bm R_j \cdot
  \nabla \times \bm W_j(\bm r-\bm R_j^o)].
\end{eqnarray}
Since $\rho, \bm{B}$ and $V, \bm{W}$ are first order quantities, the changes in
density and potential 
\footnote{The fields $\bm B$ and $\bm W$ are only defined up to a gauge
transformation: $\bm B \rightarrow \bm B + \nabla \phi$ leads to
the same physical deformation density for any scalar `potential' $\phi$.
Nevertheless, it makes sense to specify $\bm B$ and $\bm W$ uniquely (up to a
constant) by requiring it to be divergenceless.  The Helmholtz
prescription provides this divergenceless field. \cite{wep}}
can be related by linear response theory.\cite{wep}  The deformation arises solely from
off-diagonal components of the dielectric matrix $\varepsilon({\bm{q}+\bm{G},\bm{q}+\bm{G}'})$,
i.e., deviation of $\varepsilon({\bm{r},\bm{r}'})$ from the $\varepsilon({\bf r-r'})$ form.

Equation (\ref{eqn:dens}) allows for a transparent interpretation of the quantities $\rho_j$ 
and $\nabla \times \bm B_j$. The total charge density $n(\bm{r};\{\bm R_j\})$ of a 
system of \textit{displaced} atoms is constructed from the charge densities 
$\rho_j(\bm r- \bm R_j^o -\delta \bm R_j)$ that move \textit{rigidly} with the atoms 
upon displacement, plus a second part $\nabla \times \bm B_j(\bm r-\bm R_j^o)$ that 
describes how the charge density \textit{deforms} due to nuclear displacement.

It is important to keep in mind that, although the rigid enatom density (potential) is a speci\-fied decomposition of the crystal analog,
it does not arise simply from screening of the pseudopotential (which is the case in weak pseudopotential theory).
It is intrinsically a dynamically determined quantity, involving only linear response.  
Specifically, the enatom potential arises from a screened displaced (pseudo)potential 
\be
\nabla_j v = \varepsilon^{-1} \nabla_j v_{ps},
\ee
while the enatom density arises from the linear change in wave functions
\be
\nabla_j n = 2 \left( 2 Re \sum_{kn}^{occ} \psi_{kn}^* \nabla_j \psi_{kn} \right).
\ee
which can be obtained from first-order perturbation theory.
(The first factor of 2 is for spin.)

A related quantity is the atomic deformation potential $\nabla_j \epsilon_{kn}$
(change in any band energy due to displacement of the atom at $\vec R_j$),
given from perturbation theory by
\be
\nabla_j \epsilon_{kn} = \langle kn|\nabla_j v|kn \rangle
\ee
in terms of enatom quantities.  Khan and Allen showed the relation of this
{\it deformation potential} to electron-phonon matrix elements.\cite{khan}  
Resurgent interest in electron-phonon
coupled superconductivity has led to the suggestion by Moussa and Cohen that this
quantity may provide insight into strong coupling.\cite{Cohen}

Although Ball was the first to introduce the enatom decomposition and begin
to make use of it, the importance of $\nabla_j n(r)$ had been recognized earlier.
Sham emphasized its essence in the formulation of lattice dynamics, related it to the shift in potential by the density response function,\cite{sham} and tied its integral to
effective charges.  Its application to ionic  insulators was extended by Martin,
who showed that the enatom dipole and quadrupole moments are the fundamental
atomic entities that underlie piezoelectricity.\cite{martin}  

As powerful as the enatom concept is (see discussion below), very little use has
been made of it.  Falter and collaborators have adopted a related
{\it quasi-ion} idea for sublattices of multiatom compounds, and used linear
response theory (or models) to evaluate sublattice charges for
Si.\cite{falter85,falter89}
Ball and Srivastava calculated some aspects of the rigid and deformation
parts of the density in Ge and GaAs from bond-stretch distortions.\cite{BS}
No calculation of single enatom quantities yet exist for any material.


\subsection{Motivation}
The enatom concept will be particularly important in studying and understanding
phonons and electron-phonon coupling, which requires only information arising from an
infinitesimal displacement of atoms (thus, linear response).
Current implementations of linear response theory calculate the first-order change in potential due to a given phonon, and uses periodicity and Bloch's theorem to reduce the calculation to a unit cell (still time-consuming).  This linear response problem must be solved separately for each phonon momentum $\bm q$.  Using the enatom concept and linear superposition, it is necessary only to calculate the enatom density and potential {\it once} (for each inequivalent atom in the primitive cell) and perform elementary integrals necessitating only linearly superimposed, overlapping enatom potentials to calculate the 
phonon frequencies.  (We are concerned here only with metals; Ball has shown that insulators with long-range potentials require extra considerations.\cite{Ball2})  The electron-phonon
matrix elements are even easier, as they can be reduced to calculating the matrix elements of the enatom potential of each inequivalent atom only.  This might not be quite as easy as it sounds, 
because the enatom may in some cases have to be calculated out to a distance of several shells of neighbors to obtain convergence of the integrals.  
We postpone the phonon problem to future work.

\begin{table}
\begin{ruledtabular}
\begin{tabular}{c cccc cccc }
 & $P$  & $V/V_0$ & $a $ &  $n_0$ 
& $N(0)$ & $l_\mathrm{TF}$ & $E_F$ & $k_F$ \\
\hline
Li&  0  & 1.00  & 7.98 &  0.79  & 3.41   & 1.13 & 0.27 & 0.52 \\
&  35$^a$ &  0.52  & 6.41& 1.52  & 2.58 & 1.02   & 0.30 & 0.55 \\
&  50 & 0.44 & 6.05  & 1.81  &2.39  & 0.98   & 0.29 & 0.54 \\
\hline 
Al&  0 & 1.00 & 7.50 & 2.85 &2.61 & 0.91 & 0.83  & 0.91  \\
&  35 & 0.77 & 6.89 & 3.67 &1.96 & 0.87 & 0.97  & 0.98  \\
&  50 & 0.73 & 6.75 & 3.90 &1.85 & 0.86 & 1.01  & 1.00  \\
\end{tabular}
\end{ruledtabular}
\caption{\label{tab:T1}
Structural and electronic properties of fcc Li and Al as a function of
pressure.
Except for the calculated pressure ($P$), which is in GPa, all 
the quantities are expressed in atomic units.
$V_0$ is the theoretical equilibrium volume, $a$ is the fcc lattice constant, $n_0$ is the mean density of electrons in 
$10 ^{-2} el/a_B^{-3}$; $N(0)$ is the density of states at the Fermi level in
 states/(spin Ry atom),  $l_\mathrm{TF}$ is the Thomas-Fermi screening length, E$_F$ 
is the Fermi energy (the occupied bandwidth), and k$_F$ is the Fermi momentum. 
$^a$Here $a$ is the experimental lattice constant at 35 GPa; our theoretical pressure is 30 GPa.}
\end{table}

For a first detailed application of this enatom concept to enhance understanding
of bonding and electron-phonon coupling, we choose the simple metals Li and Al.  
Lithium has attracted renewed interest due to
the recent discovery that, in spite of being a simple free-electron-like metal
that is not superconducting above 100 $\mu$K at
ambient pressure,\cite{lang99} it displays high T$_c \approx$ 15-17 K in the 
30-40 GPa range,\cite{shimizu,struzhkin,schilling}
and T$_c$ = 20 K has been reported\cite{shimizu} around 50 GPa.  
This discovery made Li the best superconducting elemental metal
(now equaled by yttrium\cite{hamlin} and apparently surpassed
by calcium\cite{yabuuchi06}). 
The evolution of the electronic structure and electron-ion scattering within
the rigid muffin-tin approximation is well studied.\cite{shi,christensen}
Application of microscopic superconductivity theory, with phonon frequencies and electron-phonon (EP)
matrix elements calculated using linear response methods,\cite{profeta,deepa,deepa2,maheswari} 
has established that 
this remarkable level of T$_c$ results from strong increase of EP coupling under pressure.
The one aspect of the electron-phonon behavior in Li that is not yet understood\cite{deepa2} is 
the strong branch  
dependence of electron-phonon matrix elements.  Application of enatom techniques
promises to be an ideal way to approach the remaining questions.

Aluminum is the simplest trivalent metal, with T$_c$ = 1.2 K at ambient
pressure; superconductivity is suppressed with pressure, with T$_c < 0.1$
K at 6 GPa.\cite{Al:Tc:pressure}
Under pressure the electronic structure of Al remains that of a free
electron-like metal, and a structural transition to a hcp phase takes
place only at $P > 217$ GPa.\cite{Al:structure:pressure}
Li, on the other hand, becomes more and more covalent and undergoes several phase transitions. \cite{Li:structure} In the case of metals we use the term ``covalency'' in a loose sense to indicate the appearance of directional bonds.
The vibrational and electronic properties of the two systems
also display important differences.
While the electronic structure, Fermi surface and vibrational spectrum of
Al follow a completely normal trend (i.e., the band dispersion becomes 
steeper, the Fermi surface is virtually unchanged and the phonon spectrum
is hardened), in fcc Li 
the FS evolves from a typical s-like shpere into a multiply-connected (Cu-like) shape, with necks extending through the L points, reflecting the increase in p character.
In the phonon spectrum, structural instabilities
appear around 35 GPa along the $\Gamma-K$ line, due to the strong e-ph
coupling of some selected phonon modes, whose wave vector ${\bf q}$ connects 
the necks on the Fermi surface (cf. Refs.~\onlinecite{profeta}, \onlinecite{deepa}).
We therefore expect that also the enatom of the two systems will display 
different behaviors under pressure.


\section{Calculational approach}

\begin{figure}[tb]
\begin{center}
\includegraphics*[width=0.9\linewidth]{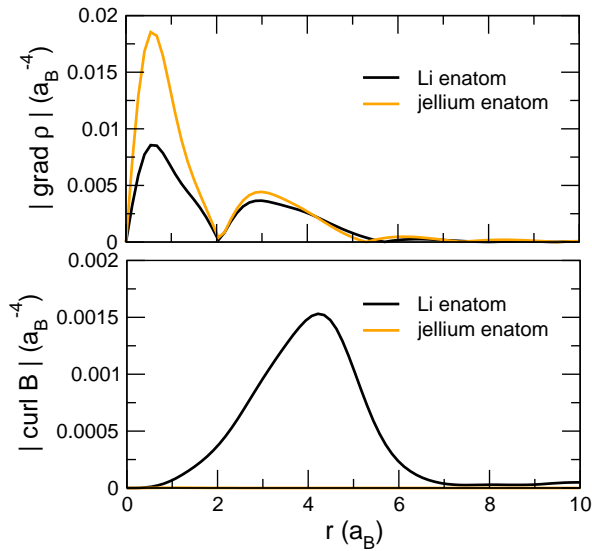}
\caption{(Color online) Comparison of the magnitude of the rigid (top) and deformation parts (bottom) in the first oder change of the density (Eq.~(\ref{eqn:foc})) of \textit{Li} at 35 GPa and a jellium model.
The plot is taken along a [110] direction through the atom.
The magnitude of $|\nabla \times \mathbf{B}|$ in jellium is
three orders of magnitude smaller than in Li, and therefore invisible on the
scale of the plot.
}
\label{fig:Li-gradcurl}
\end{center}
\end{figure}

The enatom could be computed by evaluating the linear response 
of a system to the
displacement of a single atom, that is, the dielectric response
$\epsilon(\bf r,\bm r') \leftrightarrow \epsilon(\bm q + \bm G, \bm q + \bm G')$.  
While this may be a
viable approach, it is demanding and tedious and we use another method that requires only minor 
additional codes.
Our approach is to let the computer do the linear response for us, by
choosing a supercell, displacing one atom (taken to be at the origin),
and obtaining the linear changes $\nabla_j n(\bm r)$ and $\nabla_j v(\bm r)$
by finite differences.  
Using cubic symmetry, displacement in a
single direction is sufficient to obtain the full vector changes.
The enatom components (rigid and deformation)
are then obtained using the Helmholtz decomposition of a vector field
into its irrotational and divergenceless parts. \cite{wep}

We obtained the enatom for fcc Li and fcc Al at atomic volumes corresponding to 
0, 35, and 50 GPa pressure. We used a cube supercell of lattice constant ${\cal A}$=3$a$ 
(lattice constants are listed in Table~\ref{tab:T1}), which contained 4$\times$3$^3$=108 atoms. 
The enatom is represented as a Fourier series in the supercell.
For the jellium calculations we set a single Li or Al atom into cubic supercells whose lattice constants correspond to the P=35 GPa cases. The mean electronic density was made equal to the related crystalline systems and a homogeneous positive jellium background provided charge neutrality.

For the self consistent  
density functional calculations we employed the PWSCF code \cite{PWscf} and Troullier-Martins \cite{Troullier-Martins}
norm conserving LDA pseudopotentials
and a plane wave cutoff
energy of 20 Ry for both Li and Al. 
For the {\bf k}-space integration in the primitive fcc unit cell we used a 
$(18)^3$ Monkhorst-Pack grid, \cite{MPgrid} with
a cold-smearing parameter of 0.04 Ry.\cite{Coldsmearing} 
With these parameters, we obtained a convergence of 0.2 mRy in the total energy 
and of 0.2 GPa for the pressure at 35 GPa for both systems.
For the large cubic supercell, we used a $2^3$ Monkhorst-Pack mesh, yielding four
points in the irreducible Brillouin zone, and the same cold-smearing parameter of 0.04 Ry. 
With this choice, the total energy (pressure) calculated in the
supercell equals that of the
original fcc lattice within 0.1 mRy and 0.1 GPa, respectively.
The pressure was calculated from a Birch-Murnagham fit of the LDA
$E$ vs $V$ curve. \cite{Al:syassen}$^{,}$
\footnote{To  
test the applicability of the pseudopotential method to high pressures, we 
calculated the energy vs. volume relation for the two systems and fitted it to a
Birch-Murnaghan equation, to extract the equilibrium volume ($V_0$) and the bulk modulus 
at zero pressure ($B_0$) and its derivative at zero pressure ($B_0'$).
For Lithium we obtained : V$_0$ = 127.16 $(a_\mathrm{B})^3$, B$_0$ = 14.9 GPa,
B$_0'$ = 3.33. For Al we obtained: V$_0$ = 105.3 $(a_\mathrm{B})^3$, B$_0$ = 82.7 GPa, 
B$_0'$ = 4.2, in reasonable agreement with the experimental values (V$_0$ = 111.2, 
B$_0$ = 72.7 GPa, B$_0'$ = 4.3).\cite{Al:syassen} }

In Table~\ref{tab:T1} we summarize the most relevant properties of Al and Li
as a function of pressure. Since the independent variable in our calculations is the volume of the unit cell, we also include a column showing the relative volume change.
We notice that the lattice constant of Li decreases very rapidly with pressure, 
as signaled also by the very small bulk modulus. At $P=30$ GPa, the unit cell volume of Li 
is already one half of its $P=0$ value. For comparison, the volume 
of Al at 50 GPa is 73\% of its zero pressure value.

\begin{figure}[tb]
\begin{center}
\includegraphics*[width=0.9\linewidth]{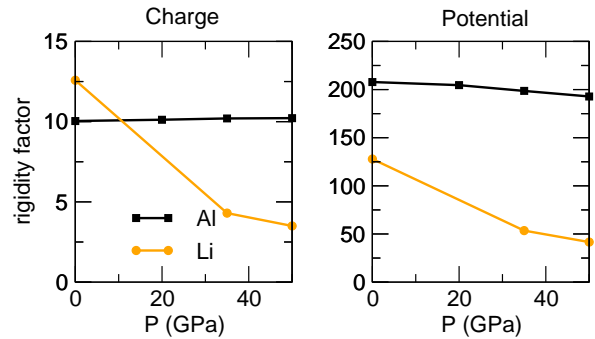}
\caption{(Color online) Evolution of the rigidity factor ${\cal R}$ (Eq.~(\ref{eq:rigidity})) of 
the enatom for Li and Al as a function of pressure.  The difference is the decrease
in rigidity with pressure in Li, which is magnified somewhat by Li's larger compressibility.}
\label{fig:rigidity}
\end{center}
\end{figure}


\section{Analysis}

As a test of our numerical approach we checked that the sum rules of 
Eqs.~(\ref{eqn:sumrule1}) and (\ref{eqn:sumrule2}) were almost perfectly fulfilled.
For the jellium enatom the deformation part should be exactly zero, as discussed in Sec.~\ref{sec:background}.
The deformation is not identically zero for our jellium enatom 
due to supercell effects.  These effects are however
very minor, {\it viz.} the maximum of the jellium deformation density is 500 times smaller than
the maximum of the Li crystal deformation density.

\begin{figure*}[tb]
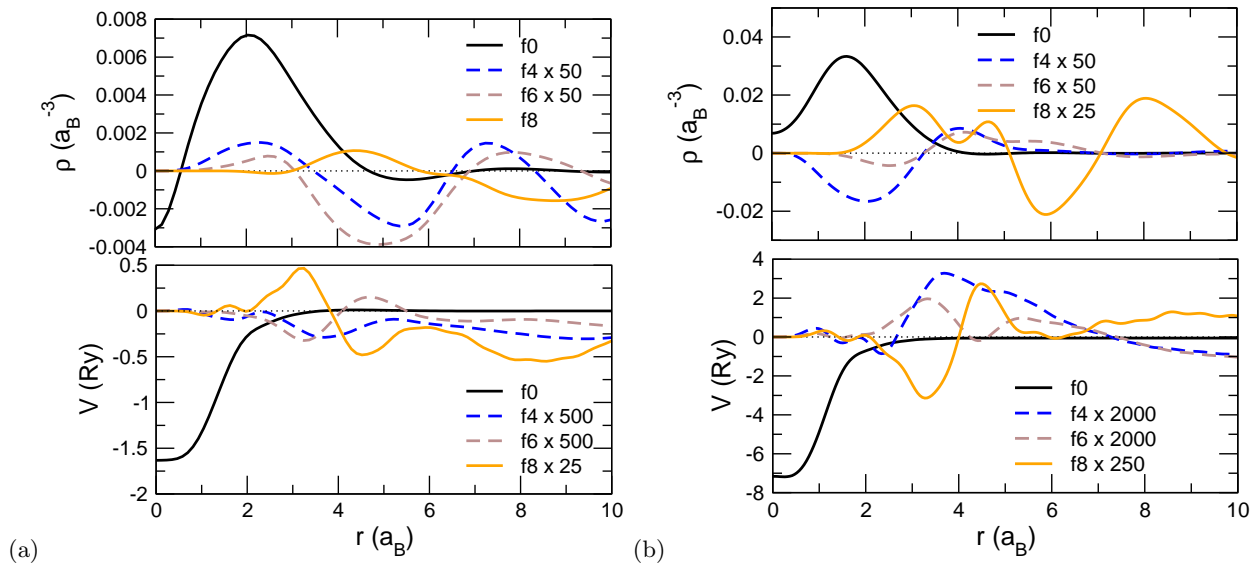

\begin{center}
\subfigure(a){\includegraphics*[width=0.9\columnwidth]{Li-decomp.eps}}
\subfigure(b){\includegraphics*[width=0.9\columnwidth]{Al-decomp.eps}}
\caption{(Color online) A kubic harmonic decomposition of the rigid parts $\rho$ and $V$ of the enatoms of (a) Li and (b) Al at 35 GPa. The radial expansion coefficients $f_L$ are defined in Eq.~(\ref{eqn:decomp}).}
\label{fig:decomp}
\end{center}
\end{figure*}

\subsection{Deformation vs rigid part}

In Fig.~\ref{fig:Li-gradcurl} we show the magnitudes of the vector fields  $\nabla \rho$ 
and $\nabla \times \bm B$ along a [110] direction. The rigid parts for fcc Li and the 
Li-in-jellium model have the same magnitude and overall shape, while the deformation parts 
are very different.  For Li $|\nabla \times {\bm B}|$ is approximately one order of magnitude smaller
than $|\nabla \rho|$.
This behavior reflects 
a general trend expected 
in simple metals: the deformation part is much smaller than the 
rigid, nevertheless it is quite revealing, as we demonstrate below. 

To quantify the strength of the fields we are considering in a more precise way, we define 
the magnitude ${\cal M}[A]$ of a scalar or vector field $A(\bm r)$ as the root mean square
\be
{\cal M}[A] = \sqrt{ \frac{1}{\Omega} \int_{\Omega} d^3\bm{r} \: [A(\bm{r})]^2},
\ee
where $\Omega$ is the volume of the supercell. 
The relative importance of $\nabla \rho$ and $\nabla \times \bm B$ in Eq.~(\ref{eqn:foc}) can be quantified by defining a \textit{rigidity factor} as
\be
{\cal R}=\frac{{\cal M}[\nabla \rho]}{{\cal M}[\nabla \times \bm B]}.
\label{eq:rigidity}
\ee
This ratio is one measure of how rigidly the enatom density (or potential, defined 
analogously) follows the nucleus.
For the perfect jellium enatom ${\cal M}[\nabla \times \bm B] = 0$ and the rigidity factor 
would diverge. But because of the already mentioned supercell effects, we
obtain ${\cal R} \sim 3000$ (1400) for the density and ${\cal R} \sim 2500$ (1500) for the
potential of jellium Li (Al), which demonstrates again that the supercell effects are indeed small.  

In the actual compounds,
for the charge, ${\cal R}$ has similar values at zero pressure but decreases by
a factor of 3 between 0 and 50 GPa in Li, whereas 
it is unchanging in Al, as shown in Fig.~\ref{fig:rigidity}.  For the potential, ${\cal R}$ is 
almost a factor of 2 smaller in Li than in Al at zero pressure, and again decreases with pressure while that for Al remains constant.
This very different behavior in two simple metals is further corroboration that Li is
increasing in covalency with pressure, while Al is not.   
The large values of ${\cal R}$ for the potential reflects the fact that the change in potential (e.g. due to phonons) is dominated by a rigid part,  which provides justification for a rigid screened ion or the rigid muffin tin potential approximation.\cite{gaspari}

\subsection{Rigid part}

\subsubsection{Lattice harmonic decomposition}

\begin{figure*}[t]
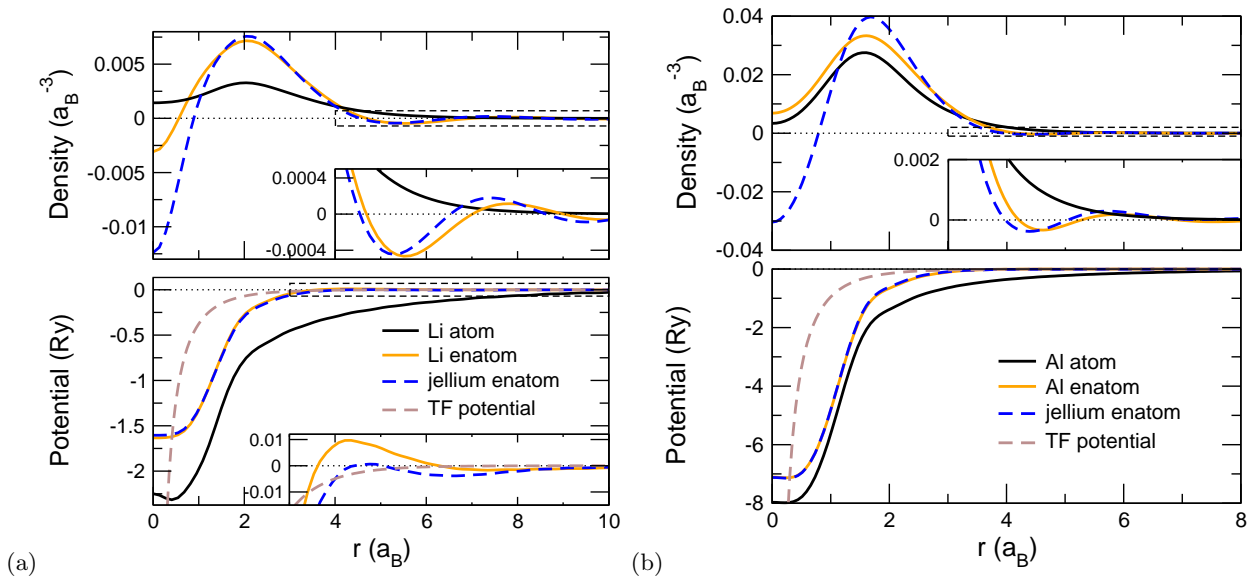

\subfigure(a){\includegraphics*[width=0.9\columnwidth]{Li-rho_jell_at.eps}}
\subfigure(b){\includegraphics*[width=0.9\columnwidth]{Al-rho_jell_at.eps}}
\caption{(Color online) Radial plots of the rigid parts of density and 
(local, $l=2$)
potential for (a) Li and (b) Al, both at 35 GPa.
We show the isolated atom, the enatom, the enatom of a jellium model,  and, only in the 
lower panels, the Thomas-Fermi potential.
The plots show the spherical parts $f_0 $ of these quantities.
On the radial direction the eight nearest neighbors are located at
(a) 4.53, (b) 4.87 $a_B$, the six next nearest neighbors are at (a) 6.41, (b) 6.89 $a_B$ and the edge of the supercell is at (a) 13.59, (b) 14.62 $a_B$, illustrating the very small overlap from neighboring supercells.
In the inset we show a blow-up of the tail region, where the Friedel
oscillations of the charge are clearly visible (see also inset in Fig.~\ref{fig:rigid-P}).}
\label{fig:rigid}
\end{figure*}

To define the degree of sphericity of the rigid density $\rho$ and potential $V$ these scalar functions can be expanded in lattice harmonics of full cubic symmetry, identified with angular 
variation $L$=0, 4, 6, 8...:
\begin{eqnarray}
\label{eqn:decomp}
\rho(\bm r) = \sum_L f_L(r) K_L(\hat{\bm r})
\end{eqnarray}
where $K_L$ is the kubic harmonic \cite{vdlage} built from spherical harmonics of angular momentum $L$ (see Ref. \onlinecite{lehmann}), $f_L$ are the radial expansion coefficients, and $r = |\bm r|$. In our case the functions $K_L$ are normalized 
according to $\int(K_L)^2 d\Omega = 4 \pi$, which ensures that the first radial expansion 
term is equivalent to the spherical average, i.e., $f_0 (r) = 1/(4\pi) \int \rho(\bm r) d\Omega$.

In Fig.~\ref{fig:decomp} we display the $L > 0$ radial expansion coefficients at 35 GPa, 
relative to the (obviously much larger) spherical part.
The ideal jellium enatom is perfectly spherical and thus contains
$f_0$ only (see Sec.~\ref{sec:background}). The decomposition of our supercell jellium model reveals small $L>0$ terms, due to small supercell effects.
The magnitudes of $f_L$ for Li and Al are comparable to the supercell effects in the jellium enatoms, as the supercell boundary is approached. Therefore we conclude that the lattice harmonic coefficients $f_4$, $f_6$, $f_8$ are meaningful only within 
a radius of $\sim{\cal A}$/3.

While the general characteristics and relative signs of the $L > 0$ terms 
are quite different in Li and Al, in most cases their maximum is only $\sim$ 1\%   or less of the 
maximum of $f_0$, for both density and potential.
The only exception is the $L=8$ lattice harmonic in the density of Li, which is surprisingly 
large around the nearest neighbor distance 4.5 $a_B$.  This anisotropy reflects the 
`cooperative' influence of neighboring atoms in determining the enatom character. The relative 
size of the $L=8$ peak grows from 8\% to 16\% of the maximum of $f_0$ from 0 to 50 GPa.
With increasing covalency in Li under pressure not only $f_8$ but all non-spherical contributions to the rigid density increase; this effect cannot be seen in Al.

For the enatom potentials, the non-spherical terms are small enough 
in fcc Li and Al at all volumes studied that the enatom potentials can be considered
effectively spherical,  as is the common assumption in simple metals.\cite{Ole1}

\subsubsection{\label{sec:screenig}Screening effects}

\begin{figure*}[tb]
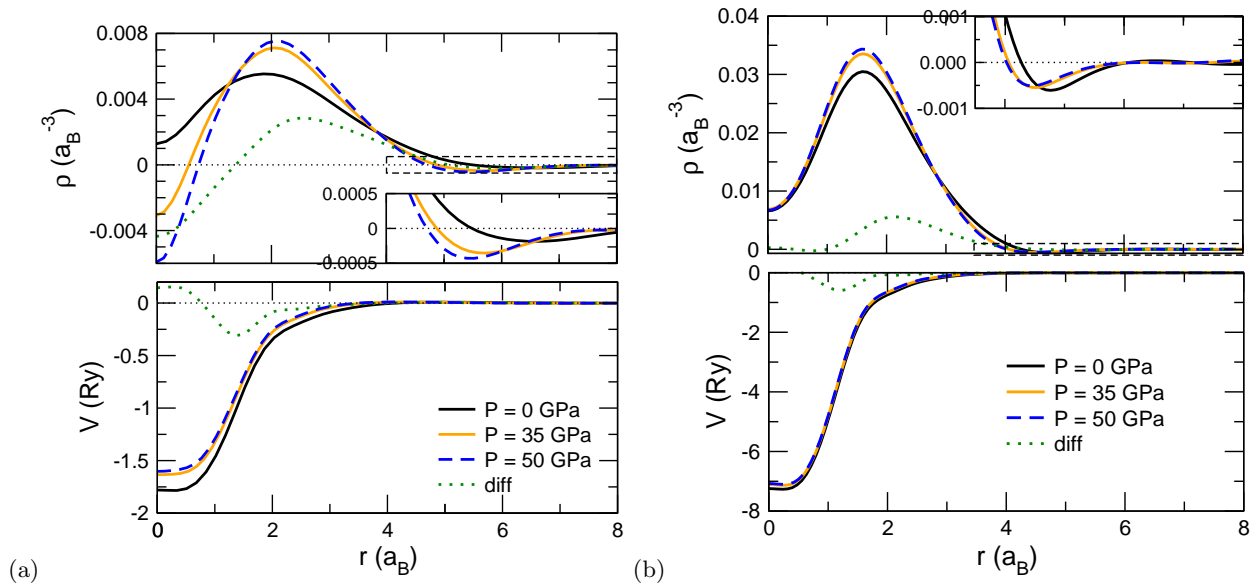

\begin{center}
\subfigure(a){\includegraphics*[width=0.9\columnwidth]{Li-pressure.eps}}
\subfigure(b){\includegraphics*[width=0.9\columnwidth]{Al-pressure.eps}}
\caption{(Color online) The pressure evolution of the rigid part of the
enatom in (a) Li and (b) Al, plotted along the [110] direction.
The dotted line represents the difference between the quantities calculated at P=35 GPa
and those at P=0.  Under pressure, the enatom density increases at its peak (at
1.5-2 $a_B$), and the local ($l=2$) potential becomes less attractive due to the increased screening. }
\label{fig:rigid-P}
\end{center}
\end{figure*}

Figures \ref{fig:rigid} and \ref{fig:rigid-P} show the rigid parts of density and potential and their pressure evolution. From linear screening theory we know the spherical part of the induced change in
charge density $\Delta n(r)$ for a simple metal 
will have long-range (but rapidly decaying) Friedel oscillations. 
Our approach reproduces the long-range oscillations according to 
\be
\Delta n(r) \sim \cos(2 k_F r)/r^3.
\ee
with reasonable agreement (see insets in Figs.~\ref{fig:rigid} and \ref{fig:rigid-P}); 
they should be important primarily for describing long-range force constants. The 
corresponding oscillations of the enatom potential are not expected to be as important, 
and we confirm this expectation, as the oscillations visible in the inset in the lower 
panel of Fig.~\ref{fig:rigid}(a) are indeed very small.

In the lower panels of Fig.~\ref{fig:rigid} we also see that the screening in the solid causes the enatom 
potentials to be significantly more short ranged than the atomic potentials. 
Furthermore, the enatom potential is less attractive by 0.7 to 0.9 Ry in these atoms.
The gradients, which determine electron-phonon matrix elements, do not seem to
differ greatly. 
Note that the pseudopotential we have used is non-local, and in the plots 
only the local ($\ell = 2$) component is shown. The total potential will include the $\ell = 0,1$ nonlocal parts,
which are non-vanishing only within the core radius ($r_{\mathrm{core}} \simeq 2$ $a_B$) and move rigidly.

To understand this screening better we compare the rigid enatom potential $V$ with the Thomas-Fermi potential
\be
V_{\mathrm{TF}}(r)=-\frac{Q^\mathrm{val}}{r}\cdot e^{-r/l_\mathrm{TF}} , 
\ee
where $Q^\mathrm{val}=eZ^\mathrm{val}$ is the total charge of the valence electrons and  $l_\mathrm{TF}$ 
is the Thomas-Fermi screening length calculated 
from the mean valence density
\footnote{Thomas-Fermi screening length: $l_\mathrm{TF} = 1/2 (\pi/3 n_0)^{1/6}$, 
$n_0$ it the mean electronic valence density in atomic units (see Table~\ref{tab:T1}).} 
(given in Table~\ref{tab:T1}). 
For both systems and all pressures we find good agreement for the long range behavior, 
confirming that the system is still dominated by homogeneous electron-gas screening. 
The electronic density of Al is higher than the one of Li and therefore the screening is stronger.
As a consequence the effective potential in Al is more localized than the one of Li.
The agreement with linear screening is better for Al than for Li, further 
supporting the deviation of Li from the homogeneous electron density picture.

\subsubsection{The rigid enatom density and potential}
The spherical average $f_0 (r)$ of the rigid enatom density $\rho$  contains a charge equal to the valence,
which is compared to a valence density for the isolated atom in Fig.~\ref{fig:rigid}, 
and also with the corresponding enatom in jellium of the appropriate density.
Note first that, while in a pseudopotential calculation the density (potential) inside 
the core radius does not have much physical meaning, changes within the core radius 
will still be useful probes of the enatom character.

The enatom density and potential are both more localized
than in the isolated atom. This difference can be ascribed to two effects. \\
(i) The density in the tail region is screened in the
solid, making the effective potential more short ranged and causing charge to move inward. 
As a result the peak value around 2 $a_B$ increases.\\
(ii) There is also charge that moves outwards from the core region, causing the peak value 
to increase further but also to move outwards.

\begin{figure}[b]
\begin{center}
\includegraphics*[width=.9\linewidth]{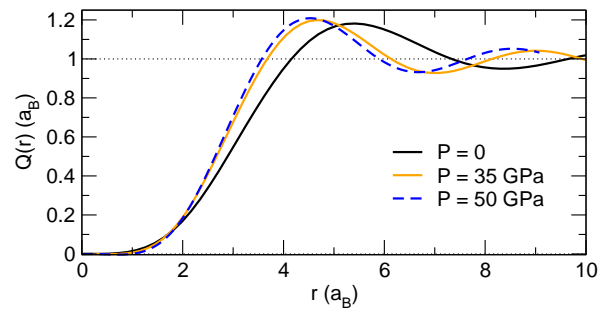}
\caption{(Color online) Integrated spherical density $Q(r) = 4 \pi \int_0^r r'^2 f_0(r') dr'$ for Li at 0, 35, and 50 GPa. The density is shown in Fig.~\ref{fig:rigid-P}.
For $r > 2$ $a_B$  the shift in density inward with pressure is evident, although there is little difference between 35 and 50 GPa. For $r < 2$ $a_B$ a small amount of charge is also shifted outward.}
\label{fig:P-int}
\end{center}
\end{figure}

The tail is very similar in jellium and in the solid (a consequence of similar Thomas-Fermi 
screening), but in the core region and around the maximum the densities are different.
The jellium enatom density becomes negative near the nucleus, with the amount of negative density  
in the core region being about 1\% of the valence, 
for both the Li and Al jellium models.
Thus this region does not contain a significant amount of negative density, but it still 
causes some charge to move away from the core. As a result the peak value of the jellium 
enatom is slightly higher and further out than the one of the crystal enatom.
(Note: in actual supercell calculations there is never any negative density, as the 
negative dip is compensated by tails of neighboring atoms.)

\subsubsection{Pressure evolution}
The pressure evolution of the rigid enatom density and potential are compared on an absolute length scale in
Fig.~\ref{fig:rigid-P}.  
Here we can identify aspects of the same two effects as described in the preceding section.
Under pressure the enhanced electronic screening makes the effective potentials more 
short ranged, and results in the screened charge moving inward (effect (i)).  
For Al the increasing pressure causes first a decrease of 
the extent of rigid density around 4 $a_B$ (see inset) and second, an increase of the peak 
value around 2 $a_B$ of about 12\% from 0 to 50 GPa. 
For Al the potential change with pressure is negligible.

Effect (i) (screening) has a stronger influence in Li than in Al because, first, the density is lower so screening is less, and second, the relative 
volume change is larger.  The peak 
value increases by more than 1/3 from 0 to 50 GPa. 

But here we also find effect (ii) which causes a small amount of charge to move away from the core and leads to an outward 
shift of the peak position  in the rigid density  from 1.9 $a_B$ (P=0) to 2.1 $a_B$ 
(50 GPa).  The shift of charge is indicated more clearly in Fig.~\ref{fig:P-int}, 
where the integrated charge is shown for Li.

The pressure evolution of the enatom potential is characterized by a decrease in
attraction at small $r$, from -1.78 Ry (P=0) to -1.60 Ry (50 GPa) in Li. This decrease 
is fairly uniform over the region out to 3.2-3.5 $a_B$ beyond which it becomes negligible.
In Al the decrease in attraction between P=0 and P=50 GPa is only 2\%.
For 35 and 50 GPa the rigid density of the Li enatom is negative inside 1 $a_B$, but the amount of 
negative density contained within that region is less than 1\%  compared to the total 
valence charge, and does not even show up in the plot of the integrated charge in
Fig.~\ref{fig:P-int}.

\begin{figure*}[tb]
\includegraphics*[width=\linewidth]{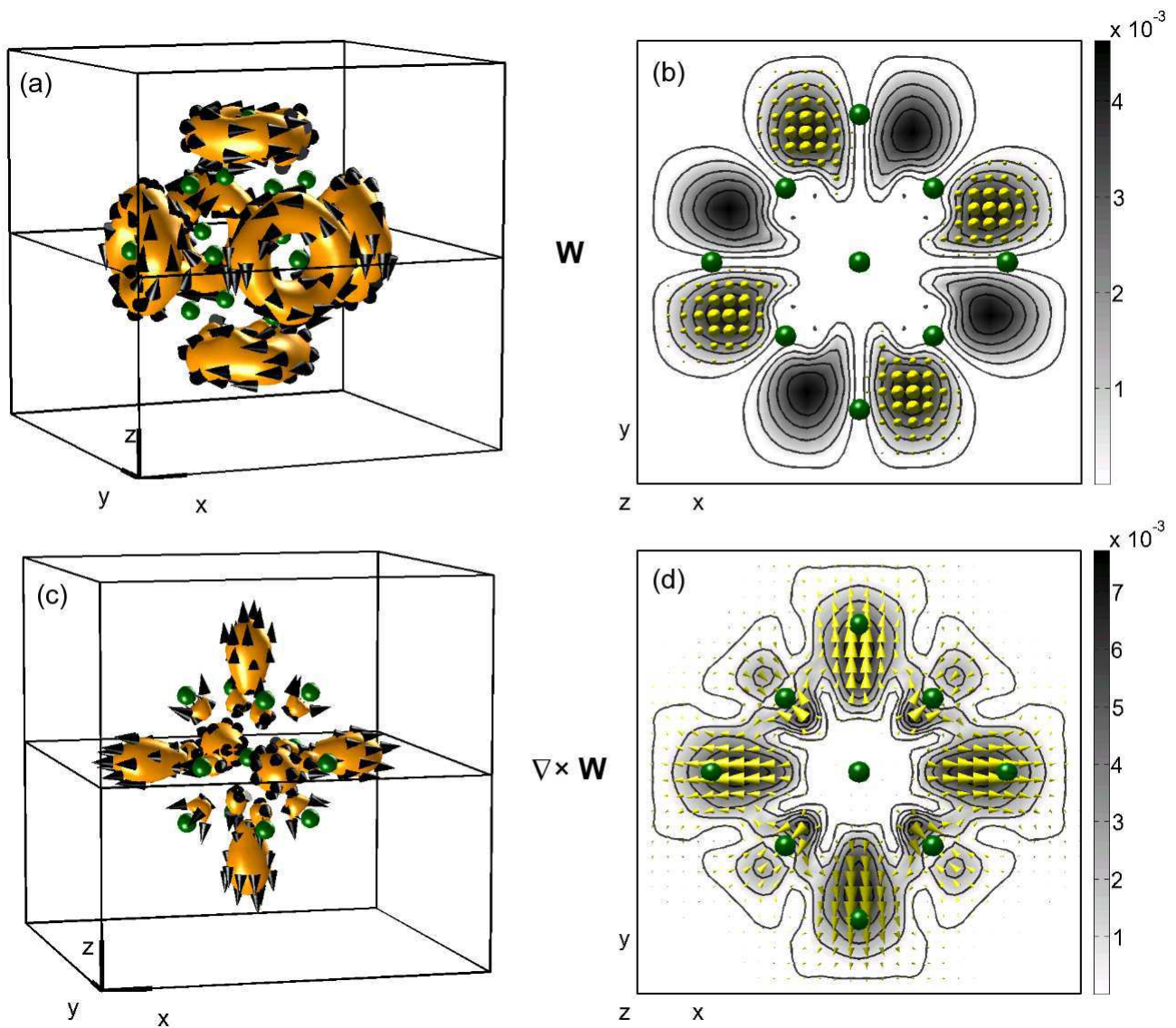}
\caption{(Color online) The deformation part of the \textit{local potential} of \textit{Li} at P = 35 GPa.
(Left-hand panels) Three dimension isocontour graphs of the magnitude, with arrows indicating the direction;
(right-hand panels) contour plots in the (001) plane.
(a) and (b) show the vector field $\bm{W}$, (c) and (d)  $\nabla \times \bm{W}$. The dark green
(dark gray) balls represent the position of the central atom and the nearest and next nearest neighbors
within the supercell, which is displayed as black boundary box.
The orange (light gray) isocontours in (a) and (c) indicate $|\bm{W}|=3.5 \times 10^{-3}$ and
$|\nabla \times \bm{W}|=4.8 \times 10^{-3}$, respectively. The black arrows are field vectors
that are located on the isocontours.
(b) and (d) indicate the magnitude of the vector fields within $xy$-planes that are indicated as
black-lined squares in the 3D graphs. Superimposed is a mesh of yellow (light gray) field vectors
which are located within the plane.}
\label{fig:3D2D}
\end{figure*}

\begin{figure*}[tb]
\includegraphics*[width=0.9\linewidth]{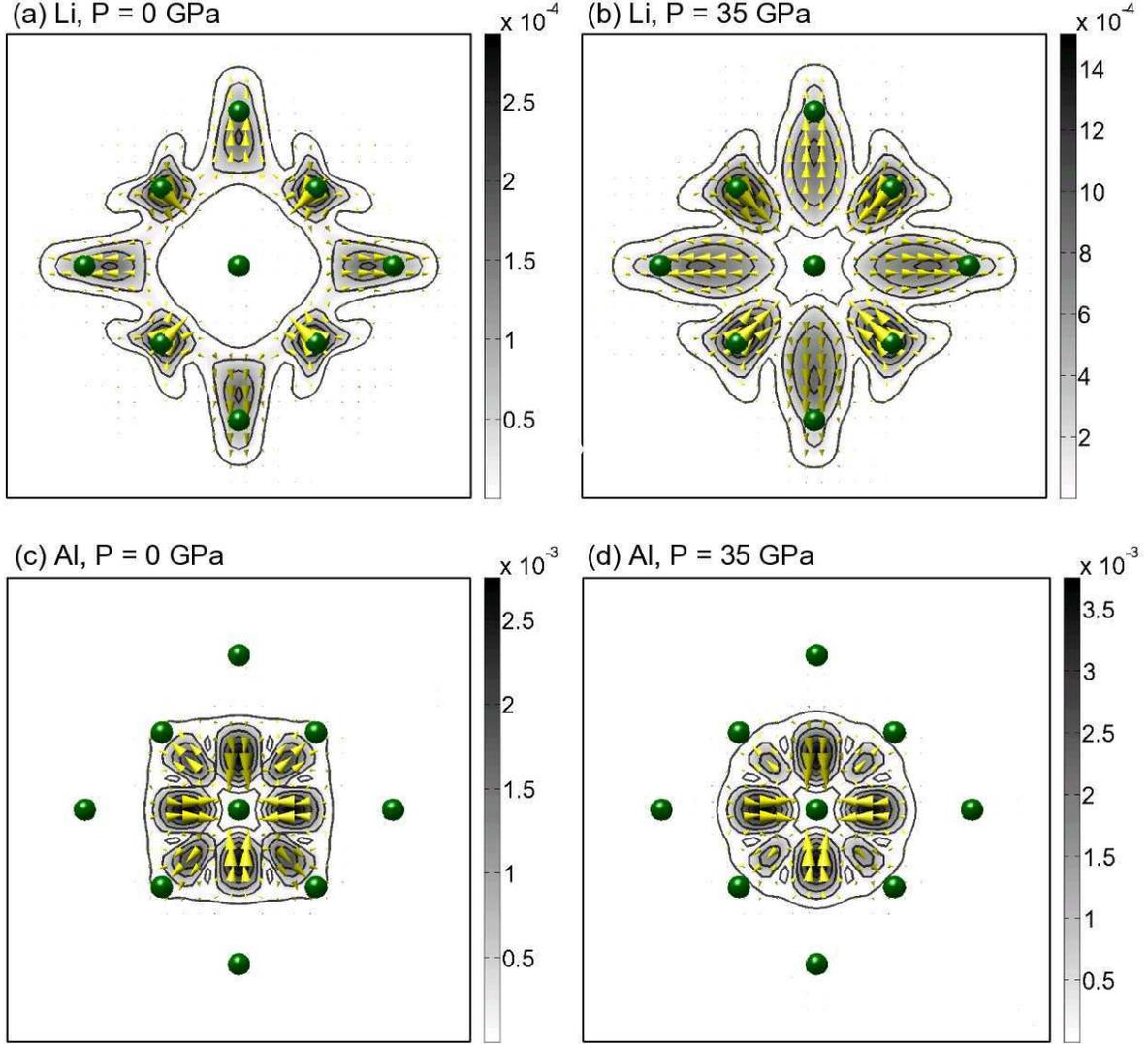}
\caption{(Color online) 2D graphs (see Fig.~\ref{fig:3D2D}) of the deformation part of the \textit{charge density} $\nabla \times \bm{B}$ for Li and Al for 0 and 35 GPa. 
Li undergoes a significant pressure evolution arising in the shape and the magnitude of $\nabla \times \bm{B}$. Al in turn changes only slightly. For the meaning of the symbols see Fig.~\ref{fig:3D2D}.}
\label{fig:deform-rho}
\end{figure*}

\begin{figure*}[tb]
\includegraphics*[width=0.9\linewidth]{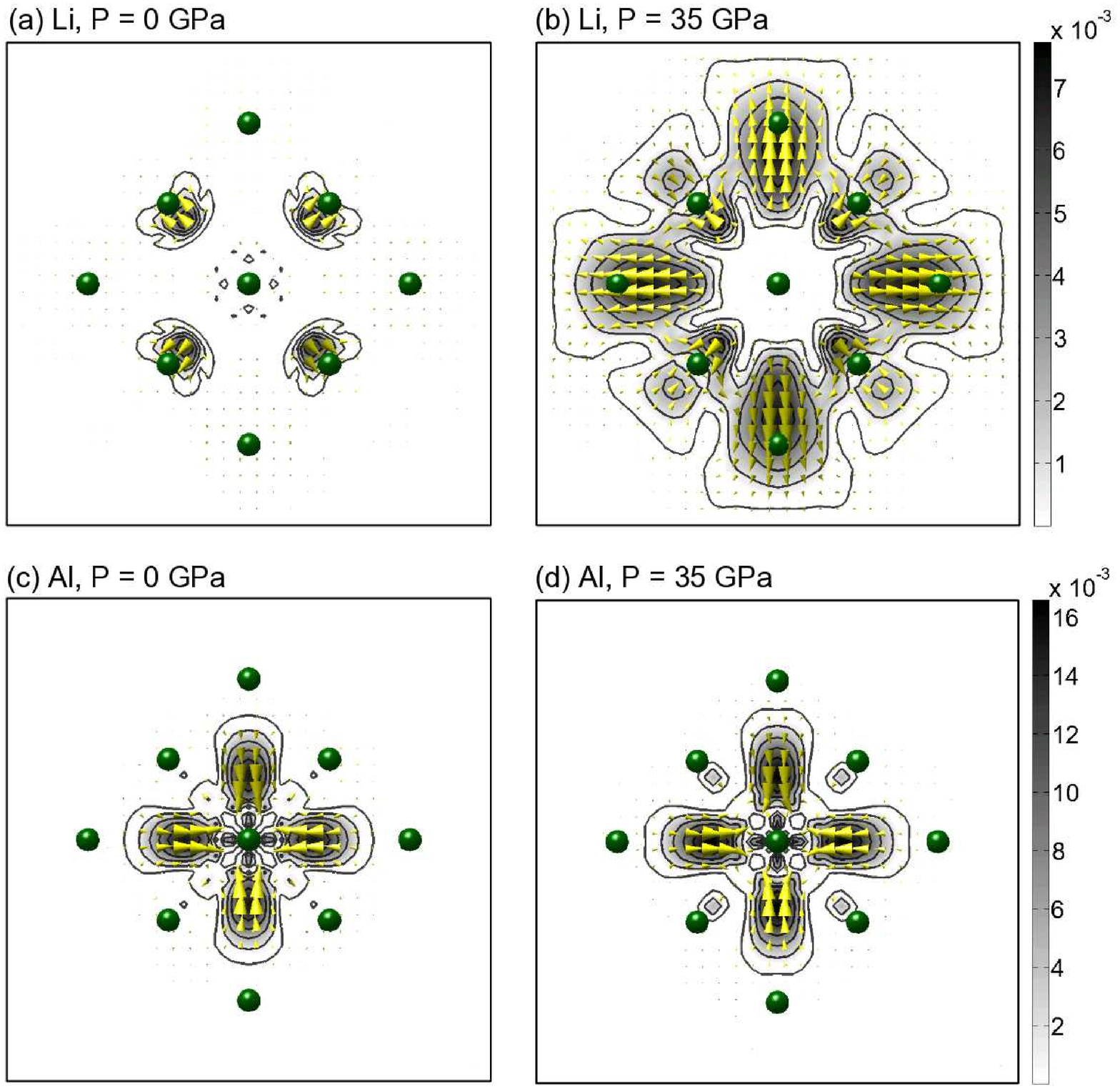}
\caption{(Color online) 2D graphs (see Fig.~\ref{fig:3D2D}) of the deformation
  part of the (local) \textit{potential} $\nabla \times \bm{W}$ for Li and Al
  for 0 and 35 GPa. Plots (a),(b) and (c),(d) share the same color bar,
  respectively. Lithium undergoes a significant change with pressure, while
  aluminum remains almost unchanged. For the meaning of the symbols see Fig.~\ref{fig:3D2D}.}
\label{fig:deform-v}
\end{figure*}

\subsection{Deformation part}           
In general the vector fields $\bm{B}$ and $\bm{W}$ (Eqs.~(\ref{eqn:dens}) and (\ref{eqn:pot})) that describe the deformation parts of density and potential 
have  similar morphologies that reflect the symmetry of the lattice. $\bm B$ or $\bm W$ 
form symmetry related ``donut''swirls centered at different distances along lines connecting the central 
atom and first (1nn) and second (2nn) nearest neighbors, i.e., the crystal axes 
(see Fig.~\ref{fig:3D2D}(a), where only swirls around the 2nn are visible). The swirls 
associated with the 1nn and 2nn have opposite rotational directions.
The derived fields $\nabla \times \bm B$ or $\nabla \times \bm W$ are large at the centers of 
the swirls of $\bm B$ or $\bm W$
(see Fig.~\ref{fig:3D2D}(c)), and they are primarily directed radially.

It is informative also to view these fields in planes as done in Fig.~\ref{fig:3D2D}(b) or Fig.~\ref{fig:3D2D}(d), where the precise position and spatial extent of their features can be judged. It can be seen, for example, that the donuts pictured in Fig.~\ref{fig:3D2D}(a) are nearly centered
on the 2nn Li sites and that $\bm W$ is oriented perpendicular to the plane, pointing either towards the viewer (vectors visible) or in the opposite direction (vectors not visible).
 As only the fields $\nabla \times \bm B$ or $\nabla \times \bm W$ are involved in the calculation of physical 
properties (see Eqs.~(\ref{eqn:foc}),~(\ref{eqn:dens}), and (\ref{eqn:pot})), we will focus our attention on them.
A comparison of the deformation parts of density and potential for two different pressures, for both Li and Al, is given in Figs.~\ref{fig:deform-rho} and \ref{fig:deform-v}. 


\subsubsection{Density deformation}

In Li at P=0 the maxima of the charge deformation $\nabla \times \bm B$,
shown in Fig.~\ref{fig:deform-rho}, are strongly localized around the nearest neighbors. Under pressure these maxima are pulled inward.  
The direction of the field determines the sign of the charge deformation.  For Li in Fig.~\ref{fig:deform-rho} $\delta {\bm R} \cdot \nabla \times {\bm B}$ (see Eq.~(\ref{eqn:dens})) is positive for $\delta {\bm R} \parallel [100]$, so there is a `charge transfer' from behind the displaced atom, to in front of it. 
Such charge distribution reflects a displacement-induced dipolar moment described by the
deformation (and which will be screened locally in a metal).
At $\pm$45$^{\circ}$ (at the 1nn sites, in fact) there is a depletion of charge, with a corresponding increase at  $\pm$135$^{\circ}$ (on the 1nn behind the displaced atom).  
In Al the pattern is similar, but the sign is reversed and the maxima
are nearer the nucleus.  These differences will affect their dynamical properties differently; this influence may be significant in Li and is probably not in Al as it remains more free-electron-like.

Under pressure, the magnitude of ${\cal M}[\nabla \times \bm B]$ in Li increases  quite significantly, being $2.8 \times 10^{-5}$  at P=0 and increasing by an order of magnitude at 50 GPa.  This change, consistent with increased covalency, is the cause of the large drop of the rigidity factor in Fig.~\ref{fig:rigidity}.  
The pressure evolution in Al is marginal: ${\cal M}[\nabla \times \bm B]$ is $1.9 \times 10^{-4}$ at P=0 and increase by only 30\% at 50 GPa.

\subsubsection{Potential deformation}
The potential deformation  in Li undergoes a surprisingly large pressure evolution, reflected in the 
shape, magnitude ${\cal M}$, and extent of $\nabla \times \bm W$. 
${\cal M}[\nabla \times \bm W]$ is $4.4 \times 10^{-4}$ at P=0 and increases by over a factor of 4 by 50 GPa. 
The contour plot of Fig.~\ref{fig:deform-v} shows the change from P = 0 to 35 GPa.  Starting with a small deformation located on the 1nn, maxima in the deformation grow in substantial regions including the 2nn. 

For Al, $\nabla \times \bm W$ has its maxima along the cubic axes, and much closer to the nucleus.  As for the charge deformation, $\nabla \times \bm W$ has the opposite sign compared to Li, and its change with pressure is minor.

Given the simple shape of the deformation term  $\nabla \times \bm W$,
it is easy to understand its effect on the total change in the potential
(Eq.~(\ref{eqn:pot})).
 $\delta {\bm R} \cdot \nabla \times {\bm W}$ 
gives an additional dipolar-type contribution, 
adding to the main change of potential $\nabla_j v$ which has a dipolar 
form arising from displacement of the (nearly spherical) rigid potential. 

The pressure evolution of the rigidity factor for the potential in Li 
(see Fig.~\ref{fig:rigidity}) shows that at 50 GPa the deformation part 
contributes about 2\% to the total change in potential $\nabla_j v$ 
(for Al this contribution is negligible).
For materials with lower rigidity the deformation part 
might give substantial contributions to $\nabla_j v$,  large enough to
affect its scattering properties or the strength of electron-phonon coupling.

Additional to the figures shown in this paper we provide several color graphs, showing examples of enatom quantities for Li in 3D and 2D views, as supplementary material for download. \cite{SuppMat}




\section{Discussion and Summary}
In this paper we have provided a numerical linear response approach, and the first 
explicit examples, of the {\it enatom} (the generalized 
pseudoatom introduced by Ball\cite{Ball1}) density and potential
for Li and Al, at pressures of 0, 35, and 50 GPa.
This enatom consists of a rigid and a deformation density (and potential). 
The rigid part defines a unique decomposition of the equilibrium density (potential) 
into atomic-like but overlapping contributions that move rigidly, to first order, with the 
nuclear position. The deformation density (potential) describes (again to first order in 
the displacement) how this charge (potential) deforms, and can be viewed as a backflow,
or (depending on its shape) as a mechanism that transfers charge from one side of the
displaced atom to the other.
The enatom quantities were obtained from supercell finite-difference calculations, 
demonstrating that this approach provides a feasible numerical treatment.

A {\it rigidity factor} $\cal R$ was introduced to quantify the relative importance 
of the rigid and deformation parts of the enatom, i.e., characterize how rigidly the 
charge (or potential) 
moves upon displacement. 
The rigidity factor is expected to be smaller for covalent materials whose bonding is 
strongly direction-dependent, and larger for metals that lack such strong bonding.\cite{Ball1}
It has been emphasized recently that Li becomes more covalent\cite{deepa,deepa2} 
under pressure; the
various components of the Li enatom have substantiated this trend and provided specific
ways in which this covalency arises.  Aluminum, on the other hand, remains quite
free-electron like up to 50 GPa.
Both behaviors are clearly reflected in the pressure evolution of the rigidity factor of 
both density and potential: it decreases by a factor of 3-4 in Li but stays almost constant in Al. 
Moreover, the rigidity of the potential is approximately one order of magnitude bigger 
than for the density. This rigidity supports the picture of a rigid potential shift with 
displacement in both Li and Al. 
In general, the rigidity factor $\cal R$ may become a useful tool for quantifying a
``generalized covalency'' of a system, even in the case of metals.

By kubic harmonic decomposition of the rigid enatom, we have shown that in Li and Al 
the potentials are effectively spherical, providing support for
spherical approximations in rigid-atom models of electron-phonon coupling.
Non-spherical contributions in the rigid density become larger as 
Li becomes more covalent under pressure, and non-rigid contributions to the potential 
increase somewhat in Li.  Changes in aluminum are much smaller.

The basic features of the spherical part of the rigid density and potential can be understood 
by means of linear screening theory (and could be calculated in that way).
First, the localization of the rigid enatom potential is a result of free electron-like 
(Thomas-Fermi) screening, showing that the mean radius of Al is smaller than the one of Li. 
Second, the tails of the rigid densities exhibit rapidly decaying Friedel oscillations.

Another finding is that the rigid enatom density is more localized than the 
density of an isolated atom. 
This is a result of two effects. 
(i) The density in the tail region is screened in the
solid, making the effective potential more short ranged and causing charge to move inward. 
(ii) There is also charge that moves outwards from the core region induced by the potential 
being less deep than in the free atom. 
The second effect is small compared to the first.
The same two effects also cause an increase of localization of the rigid enatom density when the pressure is increased.

These calculations reveal the manner in which the deformation part of the 
enatom reflects the symmetry of the lattice 
and undergoes a pressure evolution, which is quite significant in Li but small in Al. 
The basic morphological features of the deformation parts are the same in Al and in Li and 
also in the density and in the potential but their position, sign and relative magnitude is 
different. This behavior confirms the expectation that lattice symmetry 
is paramount in determining
the character of the deformation of the enatom, at least in nearly free electron
metals.

The enatom potential is exactly the quantity that determines electron-phonon matrix elements,
and it is for the electron-phonon problem that we foresee the important applications
of enatom quantities (see Sec. I.A and I.B).
The numerical procedure presented here thereby provides a viable means to an improved understanding
of electron-phonon coupling, based on a real space picture.
Furthermore, the rigidity factor $\cal R$ could be a useful tool to quantify a
``generalized covalency'' of a system, even in the case of metals.
Applications to strongly coupled elemental metals and compounds  
will be presented elsewhere.

\section{Acknowledgments}
The authors thank O. K. Andersen for ideas and encouragement.
J.K. gratefully acknowledges support from the International Max Planck Research School 
for Advanced Materials (IMPRS-AM).
W.E.P. was supported partially by National Science Foundation grant No. DMR-0421810, and 
is grateful for support from the Alexander von Humboldt Foundation.

\end{document}